\documentclass[aps,tightenlines,nofootinbib,preprint,
superscriptaddress,groupedaddress,epsfig]{revtex4}

\usepackage{graphicx}

\begin{document}

\title{The Properties of $D^{*}_{s1}(2700)^{+}$}

\vspace*{2cm}

\author{Guo-Li Wang}
\email{gl_wang@hit.edu.cn}

\author{Yue Jiang}
\email{jiangure@hit.edu.cn}

\author{Tianhong Wang}
\email{thwang.hit@gmail.com}
\author{Wan-Li Ju}
\email{scanh2000@126.com}

\affiliation{Department of Physics, Harbin Institute of
Technology, Harbin, 150001, China}

 \baselineskip=20pt

%\vskip 2cm

\begin{abstract}

\noindent

The new particle $D^{*}_{s1}(2700)^{+}$ has stimulated many
attentions. There are different assignments of its inherent
properties. It may be a $2^3S_1$, $1^3D_1$ or the mixture of
$2^3S_1-1^3D_1$ $c\bar s$ $1^-$ state. By considering its mass,
decay modes, full width, production rate, and comparing with
current experimental data, we point out that there is another more
reasonable assignment: $D^{*}_{s1}(2700)^{+}$ could be identified
as two resonances, one of which is a $2^3S_1$ state, another is a
$1D$ state, and both are $c\bar s$ $1^-$ states. The two states
have very close masses, which are around $2700$ MeV, and both have
broad decay widths. So in experiments, the overlapping of $DK$ or
$D^*K$ invariant mass distribution coming from their decays is
found, but the current experiments could not distinguish these two
resonances and reported one particle.

\end{abstract}


\maketitle

\section{Introduction}

In 2006, BABAR Collaboration first reported a broad structure
$D^{+}_{sJ}(2690)$ with mass $2688\pm 4\pm 3$ MeV and full width
$\Gamma = 112\pm 7\pm 36$ MeV \cite{BABAR}, almost at the same
time, Belle Collaboration also reported a new strange charmed
$1^-$ state $D^{*}_{s1}(2700)^{+}$, with mass $M=2715\pm
 11^{+11}_{-14}$ MeV and width $\Gamma=115\pm 20^{+36}_{-32}$
 MeV \cite{Belle1}, later modified to $M=2708\pm 9^{+11}_{-10}$ MeV
 and $\Gamma=108\pm 23^{+36}_{-31}$ MeV \cite{Belle2}.
 In 2009, BABAR showed
their new observation of $D^{*}_{s1}(2700)^{+}$ \cite{BABAR2}:
$M=2710\pm 2^{+12}_{-7}$ MeV and a broad full width $\Gamma=149\pm
7^{+39}_{-52}$ MeV. Recently, LHCb Collaboration confirmed the
existence of the $D^{*}_{s1}(2700)^{+}$ and measured its mass and
width to be $M=2709.2\pm1.9\pm4.5$ MeV,
$\Gamma=115.8\pm7.3\pm12.1$ MeV \cite{lhcb}.

Based on different models, interested theorists gave different
assignments of $D^{*}_{s1}(2700)^{+}$. In the works of Refs.
\cite{wangzg,vigande}, authors consider the possibility of it to
be a multiquark configuration, but their conclusions are negative.
Most authors believe that it is a conventional $c\bar s$ state.
For example, in Refs.
\cite{vigande,close,fazio,fazio2,zhangal,wang2700}, the first
radial excited $2~^3S_1$ ($2S$) state is favored, while in
Ref.~\cite{zhu}, the orbitally excited $1~^3D_1$ ($1D$) state is
expected, and others \cite{dmli,dmli2,zhao} believe it is most
likely a mixture of $2~^3S_1-1~^3D_1$.

Although its' existence has been confirmed by three different
experiments through different production mechanism, and the data
of its' mass and full width agree with each other's within their
error bars, $D^{*}_{s1}(2700)^{+}$ is not currently reported in
the summary table of the Particle Data Group. This shows that more
careful experimental researches and theoretical studies are still
needed. In this paper, we give another assignment of
$D^{*}_{s1}(2700)^{+}$. We claim that the current detected
$D^{*}_{s1}(2700)^{+}$ is not a single state, but could be two
overlapping states. One of them is the $2S$ dominant state with a
little $D$ wave mixed in it. Throughout the whole paper, we use
the symbol $D_{s1}^{*+}(2S)$ to label this $2S$ dominant state,
and since there may be confusion, it will be also called low-mass
state. The other is $1D$ dominant state with a little $S$ wave
mixed in it, we use the symbol $D_{s1}^{*+}(1D)$ for it, or
high-mass state if there is confusion. Because the invariant mass
distributions describing these two resonances overlap together, it
is equal to say the overlapping states are mixtures of pure $2S$
state and pure $1D$ state.

There are several reasons resulting in this overlapping assignment
of two states, we will show these in different aspects. In section
2, we give the masses and wave functions of $1^-$ $c\bar s$
states; in section 3, we show the decay modes and relative
branching ratios; the production rates of $2S$ and $1D$ states in
$B$ decays is given in section 4; finally we do some discussions
and make a conclusion in the last section.

\section{Masses and Wave Functions}

Theoretically, the $c\bar s$ $2S$ and $1D$ states have similar
masses, and the same quantum number $J^P=1^-$, so they mix
together, and become other two different states. There are many
examples, such as $\Psi(2S)$ and $\Psi(3770)$. $\Psi(3770)$ is a
$1D$ dominant state but mixed with some contribution from $2S$
wave, famous as the $2S-1D$ mixing state (there may be some
contributions from $1S$ and $3S$ states, see Ref.~\cite{richard}
for example), while $\Psi(2S)$ is the orthogonal partner of
$\Psi(3770)$, and it is $2S$ dominant state with a little $D$ wave
mixed in it. After mixing, two pure $2S$ and $1D$ states change
into another two mixing states:
$$|D^{+}_{s1}(low)\rangle=\rm cos\theta~|2^3S_1\rangle-\rm sin\theta~|1^3D_1\rangle,$$
\begin{equation}
|D^{+}_{s1}(high)\rangle=\rm sin\theta~|2^3S_1\rangle+\rm
cos\theta~|1^3D_1\rangle,\label{eq1}
\end{equation}
where the mixing angle $\theta<\pi$, and because of $S$ wave
dominant, $D^{+}_{s1}(low)$ is the low-mass mixing state, while
$D^{+}_{s1}(high)$ is $D$ wave dominant high-mass state.

The method above of mixing seems sort of artificial, so we do not
choose this method. Instead, we believe that the reason of mixing
is $2S$ and $1D$ states have the same quantum number of $J^P=1^-$,
and the forms of their wave functions are similar for the same
reason. When we choose a relativistic method based on quantum
field theory to deal with the bound state problem, the $2S$ and
$1D$ states will be obtained, and the mixing should be exist
automatically, not like man-made by hand. The Bethe-Salpeter
equation \cite{BS} is such a relativistic method to describe a
bound state, by solving it, the eigenvalue and relativistic wave
function for a bound state will be obtained. So we choose the
ordinary Cornell potential, and solve the exact instantaneous
Bethe-Salpeter equations (or the Salpeter equations \cite{salp})
for $1^{-}$ states. As expected, $2S$ dominant state mixed with a
little $D$ wave component and $1D$ dominant state mixed with a
little $S$ wave component are obtained. We will outline the key
point in this section.

The general form of a relativistic wave function for a vector
meson ($2S$ or $1D$, or their mixing states) can be written as 16
terms constructed by $P$, $q$, $\epsilon$ and the gamma matrices.
Because we make the instantaneous approximation of the BS method,
the 8 terms with $P\cdot q_{\perp}$ vanished. So the general form
for the relativistic Salpeter wave function which has the quantum
number of $J^P=1^{-}$ for a vector state can be written as
\cite{changwang,wang1}:
$$\varphi_{1^{-}}^{\lambda}(q_{\perp})=
q_{\perp}\cdot{\epsilon}^{\lambda}_{\perp}
\left[f_1(q_{\perp})+\frac{\not\!P}{M}f_2(q_{\perp})+
\frac{{\not\!q}_{\perp}}{M}f_3(q_{\perp})+\frac{{\not\!P}
{\not\!q}_{\perp}}{M^2} f_4(q_{\perp})\right]+
M{\not\!\epsilon}^{\lambda}_{\perp}f_5(q_{\perp})$$
\begin{equation}+
{\not\!\epsilon}^{\lambda}_{\perp}{\not\!P}f_6(q_{\perp})+
({\not\!q}_{\perp}{\not\!\epsilon}^{\lambda}_{\perp}-
q_{\perp}\cdot{\epsilon}^{\lambda}_{\perp})
f_7(q_{\perp})+\frac{1}{M}({\not\!P}{\not\!\epsilon}^{\lambda}_{\perp}
{\not\!q}_{\perp}-{\not\!P}q_{\perp}\cdot{\epsilon}^{\lambda}_{\perp})
f_8(q_{\perp}),\label{eq13}
\end{equation}
where the $P$, $q$ and ${\epsilon}^{\lambda}_{\perp}$ are the
momentum, relative inner momentum and polarization vector of the
vector meson, respectively; $f_i(q_{\perp})$ is a function of
$-q_{\perp}^2$, and we have used the notation
$q^{\mu}_{\perp}\equiv q^{\mu}-(P\cdot q/M^{2})P^{\mu}$ (which is
$(0,~\vec q)$ in the center of mass system).

In the method of instantaneous BS equation, the $8$ wave functions
$f_i$ are not independent. The constrain equations \cite{wang1}
result in the relations
$$f_1(q_{\perp})=\frac{\left[q_{\perp}^2 f_3(q_{\perp})+M^2f_5(q_{\perp})
\right] (m_1m_2-\omega_1\omega_2+q_{\perp}^2)}
{M(m_1+m_2)q_{\perp}^2},~~~f_7(q_{\perp})=\frac{f_5(q_{\perp})M(-m_1m_2+\omega_1\omega_2+q_{\perp}^2)}
{(m_1-m_2)q_{\perp}^2},$$
$$f_2(q_{\perp})=\frac{\left[-q_{\perp}^2 f_4(q_{\perp})+M^2f_6(q_{\perp})\right]
(m_1\omega_2-m_2\omega_1)}
{M(\omega_1+\omega_2)q_{\perp}^2},~~~f_8(q_{\perp})=\frac{f_6(q_{\perp})M(m_1\omega_2-m_2\omega_1)}
{(\omega_1-\omega_2)q_{\perp}^2}.$$

With this form of wave functions, we solved the full Salpeter
equation for $1^-$ states numerically. For strange quark, we
choose the constitute quark mass $m_s=500$ MeV; for the parameter
$V_0$, we give it by fitting the ground state mass
$M(D^*_s(2112))=2112$ MeV; for other parameters, we choose the
same values as given in Ref.~(\cite{wangzhang}). The mass spectra
of $c\bar s$ and $c \bar c$ systems are shown in Table I, where we
also show the experimental data of $c \bar c$ system from Particle
Data Group \cite{pdg}. One can see that our predictions for the
mass of $c \bar c$ $1^-$ system fit the data very well. For $c\bar
s$ vector states, we found there are two states around $2700$ MeV,
their masses are $2669$ MeV and $2737$ MeV.

To see the nature of these states, we draw the wave functions of
the first three states in Figure~1-3. The results show that, each
state has four different wave functions, $f_3$, $f_4$, $f_5$ and
$f_6$. For the first two states, the numerical values of $S$ wave
functions $f_5$ and $f_6$ are dominant, while the $D$ wave
components, $q^2f_3/M^2$ and  $q^2f_4/M^2$ ($f_3$ and $f_4$ always
show up followed by $q^2/M^2$) are neglected. So we conclude that
the first one with mass $2112$ MeV is the ground state $D^*_s$,
and the second one with a node structure is the $2S$ dominant
state. For the third state, see figure 3, $q^2f_3/M^2$,
$q^2f_4/M^2$ and $f_5$, $f_6$ are all sizable, and they look all
like $D$ waves on seemingly. But if we write them in spherical
polar coordinates, there are sizable $S$ wave components mixed in
$D$ wave (the $Y_{20}$ term of $B_{\lambda}$ in Eq. (21) in
Ref.~\cite{changwang}), so the third state is a $D$ wave dominant
state.

As for the $S$ wave dominant states, $1S$ and $2S$ states (see
Figures 1-2), we have the rough relations $f_5=-f_6$ and
$f_3=-f_4$. If we delete the negligible terms $f_3$ and $f_4$ in
Eq. (\ref{eq13}), then the vector wave function is changed to the
non-relativistic case
$\varphi_{1^{-}}^{\lambda}(q_{\perp})=(M+{\not\!P}){\not\!\epsilon}^{\lambda}_{\perp}f_5(q_{\perp})$.
But if we solve the Salpeter equation with this non-relativistic
wave function form as input, as expected, we only obtain the $1S$,
$2S$, $3S$ states, {\it et al}, and no $D$ wave states appear in
this non-relativistic case. In this case, the $D$ wave state has
to be dealt with another wave function form as input and the
mixing should be treated as the method shown in Eq. (\ref{eq1}).

\begin{table}[]\begin{center}
\caption{Our predictions for the masses (in
unit of MeV) for the $1^-$ $c\bar c$ and $c\bar s$ states, where the ground state masses of $M(D_s^*)=2112.0$
MeV and $M(J/\Psi)=3096.9$ MeV are input.} \vspace{0.5cm}
\begin{tabular}
{|c|c|c|c|c|c|}\hline
 &~$1S$ ~&~$2S$ ~&$1D$~ &~$3S$~ &~$2D$
\\\hline
${\rm Th}(c\bar c)$ &~3096.9~(input) &~3688.1~ &~3778.9~ &~4056.8~
&~4110.7~ \\\hline
${\rm Ex}(c\bar c)$ &~3096.916~ &~3686.093~ &~3772.92~ &~4040~
&~4159~
\\\hline
${\rm Th}(c\bar s)$ &~2112.0~(input) &~2669.0~ &~2737.3~ &~2994.3~
&~3033.2~
\\\hline
\end{tabular}
\end{center}
\end{table}
\begin{figure}
\centering
\includegraphics[width=0.65\textwidth]{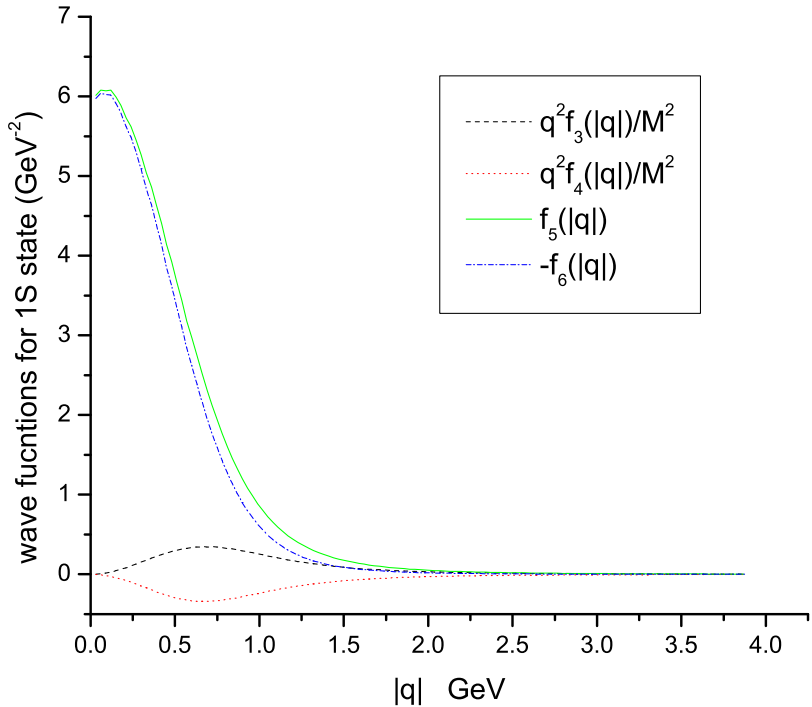}
\caption{\label{fig:1s}Wave functions for the ground state $D_{s}^{*}(1S)$.}
\end{figure}
\begin{figure}
\centering
\includegraphics[width=0.65\textwidth]{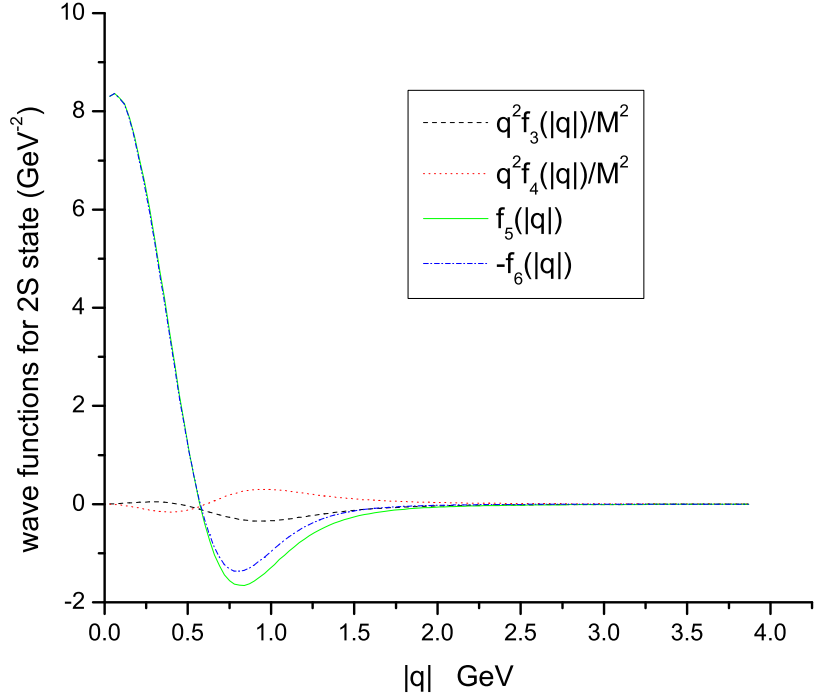}
\caption{\label{fig:2s}Wave functions for the first excited state $D_{s}^{*}(2S)$.}
\end{figure}
\begin{figure}
\centering
\includegraphics[width=0.65\textwidth]{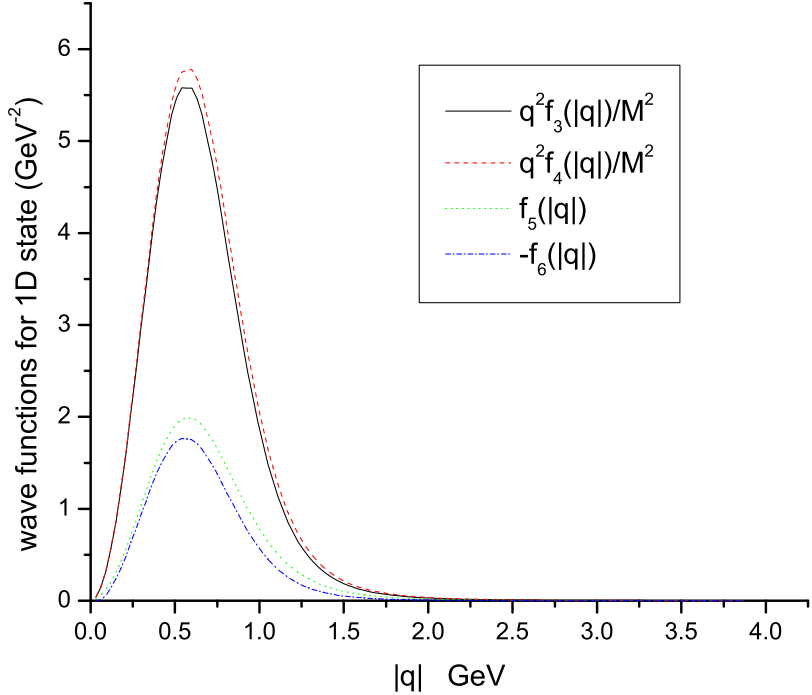}
\caption{\label{fig:1d}Wave functions for the second excited state $D_{s}^{*}(1D)$.}
\end{figure}

\begin{table}[]\begin{center}
\caption{Mass predictions for $D^{*+}_{s}(2S)$ and
$D^{*+}_{s}(1D)$ vector states in unit of MeV, where $\Delta
M=M(1^3D_1)-M(2^3S_1)$.} \vspace{0.5cm}
\label{mass}
\begin{tabular}
{|c|c|c|c|c|c|c|c|c|c|}\hline
&ours&\small{Godfrey~\cite{godfrey}}&Zeng~\cite{zeng} &
Lahde~\cite{Lahde} & Pierro~\cite{pierro} & Close~\cite{close}&
Li~\cite{dmli}&Matsuki~\cite{matsuki}&Nowak~\cite{nowak}
\\\hline
$2^3S_1$ &~$2669$ ~&~2730 ~&2730~ &~2722~ &~2806 ~&
~2711~&~2653~&~2755~&~2632~
\\\hline
$1^3D_1$ &~2737~ &~2900~ &~2820~ &~2845~ &~2913~ &~2784~
&~2775~&~2817~&~2720
\\\hline
$\Delta M$ &~68~ &~170~ &~90~ &~123~ &~107~ &~73~ &~122~&~62~&~88
\\\hline
\end{tabular}
\end{center}
\end{table}

Many authors have studied the mass spectra of $c\bar s$ $1^-$
states using different models. We list some of them which have
both the masses of $2S$ and $1D$ in Table~\ref{mass}. Except our
results, all of the predictions are pure $2S$ and $1D$ states, and
the mass shift $\Delta M=M(1^3D_1)-M(2^3S_1)$ between $2S$ and
$1D$ is also shown in Table~\ref{mass}, the values of mass shifts
lie in the region from $62$ MeV to $170$ MeV. We notice that in
the mass region of $2600\sim3000$ MeV which has been scanned by
experiments, currently there is no $1^-$ candidate except
$D^{*}_{s1}(2700)^+$. The $1^-$ candidate $D^{+}_{sJ}(2632)$
observed by SELEX Collaboration \cite{selex} has not been
confirmed by BABAR \cite{2632-b}, CLEO and FOCUS \cite{2632-c},
and its mass of $2632$ MeV seems a little small as a $2S$ or $1D$
state (see Table~\ref{mass}).

\section{Decay Modes and Relative Ratio}

Both $2S$ and $1D$ $c \bar s$ $1^-$ vector states around $2700$
MeV (or higher) can decay to $D^{0}K^{+}$, $D^{+}K^{0}$,
$D^{*0}K^{+}$, $D^{*+}K^{0}$, $D^{+}_s\eta$ and $D^{*+}_s\eta$.
Comparing with other possible decay channels, these six have
dominant branching ratios because they are all OZI allowed strong
decays. So one can use them to estimate the full widths of $2S$
and $1D$.

In a previous Letter \cite{2632}, we have already calculated the
pure $2 {^3}S_{1}$ or $1 {^3}D_{1}$ $c \bar s$ $D_s(2632)$ state
decaying to $D^{0}K^{+}$, $D^{+}K^{0}$ and $D^{+}_s\eta$
\cite{2632}, where we used the reduction formula, PCAC relation
and low energy theorem, thus the transition $S$-matrix is given as
a formula involving the light meson decay constant and the
corresponding transition matrix element between two heavy mesons.
The transition matrix element is written as an overlapping
integral of the relevant wave functions, which are obtained
numerically by solving the Salpeter equation with further
non-relativistic approximation. In this paper, we re-calculate the
OZI allowed decay modes but with full wave functions as input, and
use the new predicted masses of $2S$ and $1D$ in Table I as input,
so three more channels are opened, and the results are shown in
Table III, where the errors in our results are obtained by varying
all the input parameters simultaneously within $\pm 5\%$. We do
not repeat the calculation here, interested reader can find the
details in Ref. \cite{2632}.

One can see that, both $D^{*+}_{s}(2S)$ and $D^{*+}_{s}(1D)$ have
broad full widths, since they have at least 6 OZI allowed strong
decay channels. The full width of $D^{*+}_{s}(1D)$ is much broader
than that of $D^{*+}_{s}(2S)$, which is caused by two reasons. One
is the node structure of $2S$ wave function (see Figure 2): the
$2S$ wave function before the node provides positive contribution
to the width, while after the node it gives negative contribution.
The $D^{*+}_{s}(1D)$ state does not suffer from the node structure
since it's wave function has no node structure as a $1D$ state.
The other reason is that the higher mass state $D^{*+}_{s}(1D)$
has larger phase space. Thus the full width of $D^{*+}_{s}(1D)$ is
broader than that of $D^{*+}_{s}(2S)$.

\begin{table}[]\begin{center}
\caption{Decay widths of $2S$ dominant and $1D$ dominant states in
unit of MeV, and the last column is the summed width of these
decay channels.} \vspace{0.5cm}
\label{ourwid}
\begin{tabular}
{|c|c|c|c|c|c|c|c|}\hline &$D^{0}K^{+}$&$D^{+}K^{0}$&$D^{*0}K^{+}$
&$D^{*+}K^{0}$ &$D^{+}_s\eta$ &$D^{*+}_s\eta$ & total widths
\\\hline $D^{*+}_{s}(2S)$ &$8.9 \pm 1.2$ &$8.7 \pm 1.2$
&$12.2 \pm 1.7$ &$11.6 \pm 1.7$&$4.1 \pm 0.3$ &$0.88 \pm 0.09$ &
$46.4 \pm 6.2$
\\\hline
$D^{*+}_{s}(1D)$ &$23.3 \pm 3.2$ &$21.5 \pm 3.1$ &$12.7 \pm 1.9$
&$11.1 \pm 1.8$&$3.9 \pm 0.3$ &$0.49 \pm 0.05$ & $73.0 \pm 10.4$
\\\hline
\end{tabular}
\end{center}
\end{table}

\begin{table}[]\begin{center}
\caption{Decay widths (MeV) and their ratios with $2S$ and $1D$
assignments of $D^{*}_{s1}(2700)^{+}$ as well as the full width
(MeV), our results are from the cases of $2S$ dominant and $1D$
dominant assignments. } \vspace{0.5cm} \label{widths}
\begin{tabular}
{|c|c|c|c|c|c|}\hline
&$D^{*}_{s1}(2700)^{+}$&$DK$&$D^{*}K$&$\frac{Br(D^{+}_{sJ}\rightarrow
D^{*}K)}{Br(D^{+}_{sJ}\rightarrow DK)}$& Full width
\\\hline
ours & $2S$ & 17.6 & 23.8 & 1.35&46.4
\\\hline
Close~\cite{close}& $2S$ &22 &78 &3.55&103
\\\hline
Zhang~\cite{zhu}& $2S$ &3.2 &27.2 &8.5&32
\\\hline
Colangelo~\cite{fazio}& $2S$ & & &0.91&
\\\hline
Zhong~\cite{zhao}& $2S$ &11 &18.1 &1.65&31
\\\hline
Li~\cite{dmli2}& $2S$ &4.4 &34.9 &7.9&41.4
\\\hline
ours & $1D$&44.8 & 23.8 & 0.53&73
\\\hline
Zhang~\cite{zhu}& $1D$&49.4 &13.2 &0.27&73
\\\hline
Colangelo~\cite{fazio}&$1D$ & & &0.043&
\\\hline
Zhong~\cite{zhao}&$1D$ &148.6 & 36.3&0.24&200
\\\hline
Li~\cite{dmli2}&$1D$ & 86.8& 37.2&0.43&138.2
\\\hline
\end{tabular}
\end{center}
\end{table}

BABAR recently measured the ratios of branching fractions
\cite{BABAR2}:
\begin{equation}
\frac{Br(D^{*}_{s1}(2700)^{+}\rightarrow
D^{*}K)}{Br(D^{*}_{s1}(2700)^{+}\rightarrow DK)}=0.91\pm0.13\pm0.12.
\label{eq4}
\end{equation}
We listed theoretical predictions of these two decay widths and
their ratios with $2S$ and $1D$ assignments of
$D^{*}_{s1}(2700)^{+}$ in Table IV. One can see that in the
current existing theoretical predictions, except the Colangelo's
result, all the ratios of $2S$ assignments are larger than $1.3$,
and all the ratios of $1D$ assignments are smaller than $0.5$, so
neither the pure $2S$ nor pure $1D$ assignments of
$D^{*}_{s1}(2700)^{+}$ consist with BABAR's data. This data favor
sizable $2S$ component and sizable $1D$ component in
$D^{*}_{s1}(2700)^{+}$, so only the assignments of two overlapping
states or mixing state can match the experimental data.

\section{Production in Belle}

The Belle Collaboration \cite{Belle2} have detected the
production of $D^{*}_{s1}(2700)^{+}$ which is produced in $B^+$
exclusive decay, and the branching fraction is:
\begin{equation}Br(B^{+}\rightarrow {\bar {D}^0}D^{*}_{s1}(2700)^{+})\times
Br(D^{*}_{s1}(2700)^{+}\rightarrow
D^{0}K^{+})=(1.13^{~+0.26}_{~-0.36})\times 10^{-3}.\end{equation}

In Ref.~\cite{wang2700}, we calculated the production rate of
$D^{*}_{s1}(2700)^{+}$ in $B^+$ strong decay with two assignments,
$D^{*}_{s1}(2700)^{+}$ is a $2S$ dominant state $D^{*+}_{s}(2S)$
which is mixed with a little bit of $D$ wave component, or it is a
$1D$ dominant state $D^{*+}_{s}(1D)$ with small $S$ wave mixed in
it. With these assignments, we obtained
\begin{eqnarray}
Br(B^{+}\rightarrow {\bar {D}^0}D^{*+}_{s}(2S))=(0.72\pm0.12)\%,
\label{Swave}
\end{eqnarray}
\begin{eqnarray}
Br(B^{+}\rightarrow {\bar {D}^0}D^{*+}_{s}(1D))=(0.027\pm0.007)\%.
\label{Dwave}
\end{eqnarray}
These results are exactly proportional to the square of decay
constant of $D^{*+}_{s}(2S)$ or $D^{*+}_{s}(1D)$, so the results
are very sensitive to the values of decay constants. We know that
usually the decay constant of $2S$ state is much larger than that
of $1D$ state.  With further calculations we gave the product of
ratios \cite{wang2700}
\begin{equation}Br(B^{+}\rightarrow {\bar
{D}^0}D^{*+}_{s}(2S))\times Br(D^{*+}_{s}(2S)\rightarrow
D^{0}K^{+})= (1.4\pm 0.5)\times 10^{-3}\end{equation} and
\begin{equation}Br(B^{+}\rightarrow {\bar
{D}^0}D^{*+}_{s}(1D))\times Br(D^{*+}_{s}(1D)\rightarrow
D^{0}K^{+})= (0.9\pm0.3)\times 10^{-4}.\end{equation} The former
is consistent with Belle's data, and the later is about one order
smaller. These results show that the production of $2S$ state
$D^{*+}_{s}(2S)$ is favored, while $1D$ state $D^{*+}_{s}(1D)$ is
suppressed in the experiment of Belle.

The new state $D^{*}_{s1}(2700)^{+}$ detected by Belle is based on
an analysis of $B\bar B$ events collected at the $\Upsilon(4S)$
resonance. While in the experiment of BABAR, their analysis of
$D^{*}_{s1}(2700)^{+}$ is based on data sample recorded at the
$\Upsilon(4S)$ resonance and $40$ MeV below the resonance, and the
background from $e^{+}e^{-}\to B\bar B$ events is removed by
requiring the center of mass momentum $p^*$ of the $DK$ or $D^*K$
system to be greater than $3.3$ GeV. Unlike Belle or BABAR, LHCb
Collaboration try to made sure that the candidates of
$D^{*}_{s1}(2700)^{+}$ are produced in the primary $pp$
interaction, and reduces the contribution from particles
originating from $b-$hadron decays. It is not clear to us whether
the $1D$ state $D^{*+}_{s}(1D)$ is suppressed or not in the
experiments of BABAR and LHCb, but according to the broad full
width $\Gamma=149\pm 7^{+39}_{-52}$ MeV of $D^{*}_{s1}(2700)^{+}$
and the ratio shown in Eq.~(\ref{eq4}) detected by BABAR, it seems
that there is no suppression of $1D$ state $D^{*+}_{s}(1D)$. We
also point out that it is crucial to detect the ratio
$\frac{Br(D^{*}_{s1}(2700)^{+}\rightarrow
D^{*}K)}{Br(D^{*}_{s1}(2700)^{+}\rightarrow DK)}$ in Belle and
LHCb to see if the $1D$ state is suppressed or not.

\section{Discussion and Conclusion}

It is known that there exist the $2S$ and $1D$ vector states, and
their masses are usually close to each other. In the case of
$c\bar s$ bound states, the mass predictions by different
theoretical models are listed in Table~\ref{mass}, and we did find
that most of the predicted mass shifts $M(1D)-M(2S)$ are small. In
Table~\ref{ourwid} and Table~\ref{widths}, theoretical models also
show that both $2S$ and $1D$ are broad states. Two states whose
full widths are all broad with closed masses means that we will
find overlapping in $DK$ or $D^*K$ invariant mass distributions in
the mass regions of $2S$ and $1D$ $c\bar s$ states. Some models,
for example, Godfrey \cite{godfrey} estimated a higher mass of
$1D$ state with large mass shift $M(1D)-M(2S)$, but we argue that
higher $1D$ mass will result in broader full width. And currently,
the scanned results in experiment show that there is no other
$1^-$ state candidate in the mass region from $2600$ MeV to $3000$
MeV. One may argue that the production of $1D$ state may be
suppressed like in Belle, but the $3^-$ state of $2860$ MeV is
already found, there is no reason to believe that it is more
difficult to produce a $1D$ state than to produce a $3^-$ state.

For the pure $2S$ state assignment of $D^{*}_{s1}(2700)^{+}$,
theoretically, except for Ref.~\cite{close}, which gives a large
$2S$ full width of $103$ MeV, all the predictions show that the
pure $2S$ whose full width ranges from $31$ to $46$ MeV is too
narrow to fit data, and the ratios of $D^{*}_{s1}(2700)^{+}$
decays to $D^*K$ over those to $DK$ can not match data except the
result of Colangelo \cite{fazio}. For a pure $1D$ assignment,
though the predicted full widths can match the data, we still can
not explain the ratio of $\frac{Br(D^{+}_{s1}\rightarrow
D^{*}K)}{Br(D^{+}_{s1}\rightarrow DK)}$ detected by BABAR, and we
can not explain why we found the $1D$ state, but the $2S$ state is
not found since the later is favored while the former is
suppressed in Belle. So the pure $2S$ and pure $1D$ assignment of
$D^{*}_{s1}(2700)^{+}$ can be ruled out.

The other possibility is that $D^{*}_{s1}(2700)^{+}$ is a $2S-1D$
mixing state. As shown in the BS relativistic method, the reason
of mixing is that they have the same quantum number of $1^-$, and
the way of mixing may not follow the the method shown in
Eq.~(\ref{eq1}). Even we believe that it is correct, {\it i.e},
two states $2S$ and $1D$ mix together, and after mixing, they turn
into another two mixing states. Experiment find one of them, the
open question is, where is the orthogonal partner of this found
mixing state? One may argue that the missing one may be far away
from the region of $2700$ MeV. But if it is true, then there is no
reason to mix together.

So the most reasonable assignment is that $D^{*}_{s1}(2700)^{+}$
is not one single state, but two overlapping states,
$D_{s1}^{*+}(2S)$ and $D_{s1}^{*+}(1D)$, both of them are broad
states and there is a narrow mass gap between them. So the $DK$ or
$D^*K$ invariant mass distributions from their decays overlap
together, and then one single state $D^{*}_{s1}(2700)^{+}$ with a
broader full width is detected and reported in experiments. We
point out that, with same quantum number $J^P=1^-$ and close
masses, the two broad overlapping states can not be easily
distinguished by angular analysis in experiments. By this two
overlapping states assignment, the experimental data can be
explained easily. If we set the possibilities of production of
$D_{s1}^{*+}(2S)$ and $D_{s1}^{*+}(1D)$ are same, we roughly
estimate that a state with mass $M=2703$ MeV and full width
$\Gamma=127.7\pm8.3$ MeV will be detected by experiments, and the
predicted ratio is $\frac{Br(D^{+}_{s1}\rightarrow
D^{*}K)}{Br(D^{+}_{s1}\rightarrow DK)}=0.76^{+0.25}_{-0.19}$, all
these values consist with the experimental data.

In summary, from a study of relativistic BS method, we find around
mass region $2700$ MeV, there are two $c\bar s$ states,
$D_{s1}^{*+}(2S)$ and $D_{s1}^{*+}(1D)$, which have the same
quantum number $J^P=1^-$ and similar masses. Theoretical
calculations show that both of them have broad full width, so they
overlap together. This two states assignment of
$D^{*}_{s1}(2700)^{+}$ can fit data very well.

\noindent {\Large \bf Acknowledgements}
This work was supported in part by the National Natural Science
Foundation of China (NSFC) under Grant No.~11175051.

\appendix{}

\end{document}